\documentclass[a4paper,11pt]{article}
\usepackage{pos}
\usepackage{tikz}
\usepackage{color}
\usepackage[dvipsnames]{xcolor}
\usepackage{enumitem}
\usepackage{cleveref}
\title{Dark matter as higher-form fields }
\author*[a]{Cypris Plantier}

\affiliation[a]{Laboratoire de Physique Subatomique et de Cosmologie,\\
  Univeristé Grenoble Alpes, CNRS/IN2P3, Grenoble INP, 38000 Grenoble, France}

\emailAdd{cypris.plantier@lpsc.in2p3.fr}

\abstract{Among the crowd of dark matter candidates, light scalar and vector particles have been the focus of intense theoretical and experimental investigations over the past 15 years. However, those are always described as scalar or vector fields. Instead, we investigate their embedding as antisymmetric rank-three and rank-two tensor fields. By reconciling theoretical investigations, regarding the nature of the dualities relating the standard and tensorial descriptions and their different possible Stueckelberg gauge representations, and phenomenological examinations considering the effective interactions between tensor fields and the Standard Model, we prove these tensor fields to provide a completely distinct framework. This alternative set allows for different couplings and therefore different experimental signatures for the detection or production of dark matter at both low energy and colliders.}

\FullConference{43rd International Conference on High Energy Physics (ICHEP 2026)\\
30 July  to 5 August , 2026\\
Natal, Brazil\\}

\begin{document}
\maketitle

\section{Introduction}

 The quest for dark matter continues to motivate the exploration of new candidates, ideally both well-motivated theoretically and phenomenologically distinctive. Here, we propose to investigate two tensorial gauge-bosons as dark matter candidates: the antisymmetric two-index field $B^{\mu\nu}$ (often called \textit{Kalb-Ramond} field) \cite{Kalb:1974yc}, and three-index field $C^{\mu\nu\rho}$ \cite{Curtright:1980yk}. When massless, these higher-form fields possess an Abelian gauge symmetry, under which they transform as
 \begin{subequations}
     \begin{align}
         B^{\mu\nu}&\longrightarrow B^{\mu\nu}+\partial^{\mu}\Lambda^{\nu}(x)-\partial^{\nu}\Lambda^{\mu}(x),\;\;\;\;\;\;\;\;\;\;\;\;\;\;\;\;\;\;\;\;\;\;\;\;\;\;\;\Lambda^{\mu}(x)\text{ $1$-form gauge parameter},\\
         C^{\mu\nu\rho}&\longrightarrow C^{\mu\nu\rho}+\partial^{\mu}\Lambda^{\nu\rho}(x)+\partial^{\nu}\Lambda^{\rho\mu}(x)+\partial^{\rho}\Lambda^{\mu\nu}(x),\;\;\;\;\Lambda^{\mu\nu}(x)\text{ $2$-form gauge parameter},
     \end{align}
 \end{subequations}
and that leaves their strength-tensors $F_B^{\mu\nu\rho}$ and $F_C^{\mu\nu\rho\sigma}$ invariant. The existence of these internal symmetries, along with their peculiar Lorentz structure, makes the higher-form fields particularly interesting building blocks to design effective theories.

Higher-form fields constitute alternative representations of the usual spin-$0$ and spin-$1$ particles. Historically, the belief that these representations are strictly equivalent to the usual ones is rooted in the existence of an algebraic relation linking them, the so-called \textit{duality} \cite{Hell:2021wzm}. However, as we will show in details further, this equivalence is only valid when we compare free theories. Following how duality crumbles in the presence of external fields will lead us to several interesting phenomenological implications. At the end, we prove that higher-forms provide  completely unique frameworks whenever it comes to describing interacting dark matter. This paper constitutes an overview of some of the main findings of one of our previous work \cite{Plantier:2025hcm}.
\section{Degrees of freedom and Stueckelberg decomposition}

The numbers of independent degrees of freedom (DoFs) propagated by each free form-field in both massive and massless cases are summarized in \cref{tab:placeholder}. It highlights some crucial redundancies in the number of DoFs that can be interpreted as a freedom to describe a particle using its usual or its higher-form representation. Specifically, a massless scalar particle can be represented either by a massless $0$ or $2$-form field, a massive scalar particle by a massive $0$ or $3$-form field and a massive vector particle by a massive $1$ or $2$-form field.

Truly understanding to what extent these representations can be considered as equivalent requires the introduction of the concept of duality, that will be the subject of the two following sections. Still, it is already possible to state a strong argument about this would-be equivalence, only by regarding the way the DoFs can be decomposed using Stueckelberg representation \cite{Stueckelberg:1938hvi}.

A massive vector has two transverse and one longitudinal polarizations. Under the Stueckelberg (or the Goldstone) representation, the transverse polarizations are associated to a massless vector field while the longitudinal polarization is associated to a would-be scalar field $\chi$, thereby rendering the decomposition
\begin{equation}
    A^{\mu}_{\text{massive}}=A^{\mu}_{\text{massless}}+\frac{\partial_{\mu}\chi}{m},
\end{equation}
with $m$ the mass of the vector field. This representation is famous for  being at the core of the Higgs mechanism, as it describes exactly how the $W$ boson acquires its mass. In the Fourier space, the derivative converts to an impulsion factor that will ensure the predominance of the scalar component at high-energy: this is the Goldstone equivalence theorem. As it is well-known, a vector field is essentially its longitudinal polarization at high energy.

A similar decomposition can be found for $B^{\mu\nu}$ and $C^{\mu\nu\rho}$,
\begin{equation}
B^{\mu\nu}_{\text{massive}}=B^{\mu\nu}_{\text{massless}}+\frac{F^{\mu\nu}_A}{m},\;\;\;\;\;\;\;\;\;\;\;\;\;\;\;\;\;\;
C^{\mu\nu\rho}_{\text{massive}}=C^{\mu\nu\rho}_{\text{massless}}+\frac{F^{\mu\nu\rho}_B}{m}.
\end{equation}
A massive $2$-form field thus essentially behaves as a transverse vector field at high energy. Kinematically speaking, it is in the internal repartition of their DoFs that the massive $1$ and $2$-form stand from each other: at high energy, a $1$-form is essentially its scalar-like, longitudinal component whereas a $2$-form is essentially its photon-like, transverses components. Concerning the $3$-form field, all its dynamic is contained in its $2$-form-like component, that is itself dual to a massless scalar field. Even though dynamically its massless component cannot propagate anything, one has to note that it can still be responsible for many interesting topological effects \cite{dvali2022strongcpgravity}.

\section{Duality for free fields}\label{sec:duality}
Formally speaking, two free $p$-forms fields $\omega_1$ and $\omega_2$ propagating the same number of degrees of freedom are said to be \textit{dual} to each other if there exists a \textit{parent} Lagrangian  
$L(\omega_1,\omega_2)$ containing both fields and such as integrating out one one the fields through its equations of motion renders back the Lagrangian of describing the free propagation of the other \cite{Hell:2021wzm,Polchinski:1998rr}. Duality is a statement of equivalence: two dual fields describe the same physical phenomena.

Unsurprisingly, all the coincidences in DoFs presented in the previous section turn out to be manifestations of an underlying duality (for details concerning the form of the parent Lagrangians, see \cite{Plantier:2025hcm}). As a consequence, the usual and the higher-form descriptions of a free particle turn out to be rigorously identical.

This statement, however, does not say much about the situation where the couplings with external fields are allowed. Indeed, the fact that the rich Lorentz structure of higher-form fields opens the gate to a host of exotic couplings with matter, along with the unique kinematic behavior of higher-form fields at high-energy constitute strong hints of a breaking of the duality by external couplings. Proving this intuition in the next section will first require to properly characterize dualization in the presence of effective couplings. In the following, we will content in treating the case of massive dualities, that is the case of interest for dark-matter phenomenology.

\section{Duality and effective operators}
In the process of integrating one of the forms, dualization provides through the equations of motion an algebraic prescription to go from a field to another. For exemple, in the case of the massive $\phi$/$C$ duality, we use 
\begin{equation}\label{duality}
    C^{\mu\nu\rho}\rightarrow-\frac{1}{m}\varepsilon^{\mu\nu\rho\sigma}(\partial_{\sigma}\phi),\;\;\;\;\;\;\;\;\;\;\;\;\;\;\;\;\;\; \phi\rightarrow\frac{1}{4!m}\varepsilon_{\mu\nu\rho\sigma}F_C^{\mu\nu\rho\sigma},
\end{equation}
to dualize the $3$-form to a $0$-form and reciprocally. 

In the presence of interactions with external fields, these relations will provide a mapping between the effective operators of both basis. Schematically, for $\omega_1$ and $\omega_2$ two dual fields and $J$ a generic external current, 
\begin{equation}
    L_{free}(\omega_1)+L_{int}(\omega_1,J)\xrightarrow[]{\text{dualization}} L_{free}(\omega_2)+L_{int}(\omega_2,J).
\end{equation}
The existence of this one-to-one correspondence is in itself remarkable: for exemple, it ensures that, despite its rich Lorentz structure, $C^{\mu\nu\rho}$ will have exactly the same number of effective interactions with a given external field as $\phi$ once all the redundant operators have been eliminated.

Let us now describe in more details the properties of this mapping, sticking to the exemple of $\phi$/$C$ duality. Looking at \eqref{duality}, we observe that dualization systemically involves a mass scale and a Levi-Civita tensor, and that it relates the field to the strength-tensor of its dual. Consequently, dualization does not preserve the mass power counting, the gauge properties and the parity of the initial operator. One can convince oneself looking at the way the dominant effective coupling between $\phi$ and fermions is dualized in the $C$-basis
\begin{equation}
    L_{int}(\phi,\psi)=\phi\bar{\psi}_L\psi_R\xrightarrow[]{\text{dualization}}L_{int}(C,\psi)=\frac{1}{4!m}\varepsilon_{\mu\nu\rho\sigma}F_C^{\mu\nu\rho\sigma}\bar{\psi}_R\psi_L.
\end{equation}
Here, a renormalizable, shift-symmetry breaking, parity-even operator is mapped on a dimension-five, gauge-invariant, parity-odd operator. This statement is equally verified for each massive duality presented in \cref{sec:duality}.

Although it has be shown that two dual operators can to some extent lead to common observables (see \cite{Plantier:2025hcm} for details about duality at the squared amplitude level), the shuffling of the properties listed above upon dualization prevents us from declaring them equivalent. From the effective viewpoint, it is evident that describing an interaction using a renormalizable or a suppressed operator does not indicates the same physical frameworks, specifically concerning UV-completions. Regarding the symmetry aspect, imposing gauge-invariance to the Lagrangian is sufficient to discriminate two dual operators.

Two very useful corollary follow: firstly, two dominant effective operators cannot be dual to each other and thus nothing prevents them from having very different structures. For exemple, the dominant effective coupling of $\phi$ with fermions is a Yukawa coupling $\phi\bar{\psi}_L\psi_R$ whereas the three-form couples preferentially to  vector currents through $\varepsilon_{\mu\nu\rho\sigma}C^{\mu\nu\rho}\bar{\psi}_L\gamma^{\sigma}\psi_L$. Secondly, two gauge-invariant operators cannot be dual as well, meaning that the gauge-invariant operators of both basis can really stand from each other. The consequences for dark matter model-building will be explored in the following section.

\begin{figure}[]
    \centering  
    \begin{minipage}[t]{0.45\textwidth}
        \centering
        \vspace{0.5cm}
         \begin{tabular}{c|c|c|c}
           $p$ & $p$-form field & massless & massive \\
         \hline
         $0$ & $\phi $ & $\color{blue}1$ & $\color{red}1$\\
         $1$ & $A^{\mu} $ & $2$ & $\color{Green}3$\\
         $2$ & $B^{\mu\nu}$&$\color{blue}1$ & $\color{Green}3$\\
         $3$ & $C^{\mu\nu\rho}$&$0$ & $\color{red}1$\\
    \end{tabular}
    \vspace{1.2cm}
     \captionof{table}{Numbers of DoFs propagated by each free $p$-form field ($p$ being the number of Lorentz indices) in both massive and massless cases. Identical colors indicate a coincidence in the number of propagated DoFs.}
    \label{tab:placeholder}
    \end{minipage}
    \hspace{0.3cm}
    \begin{minipage}[t]{0.45\textwidth}
        \vspace{0pt}
        \centering
        \includegraphics[width=\linewidth]{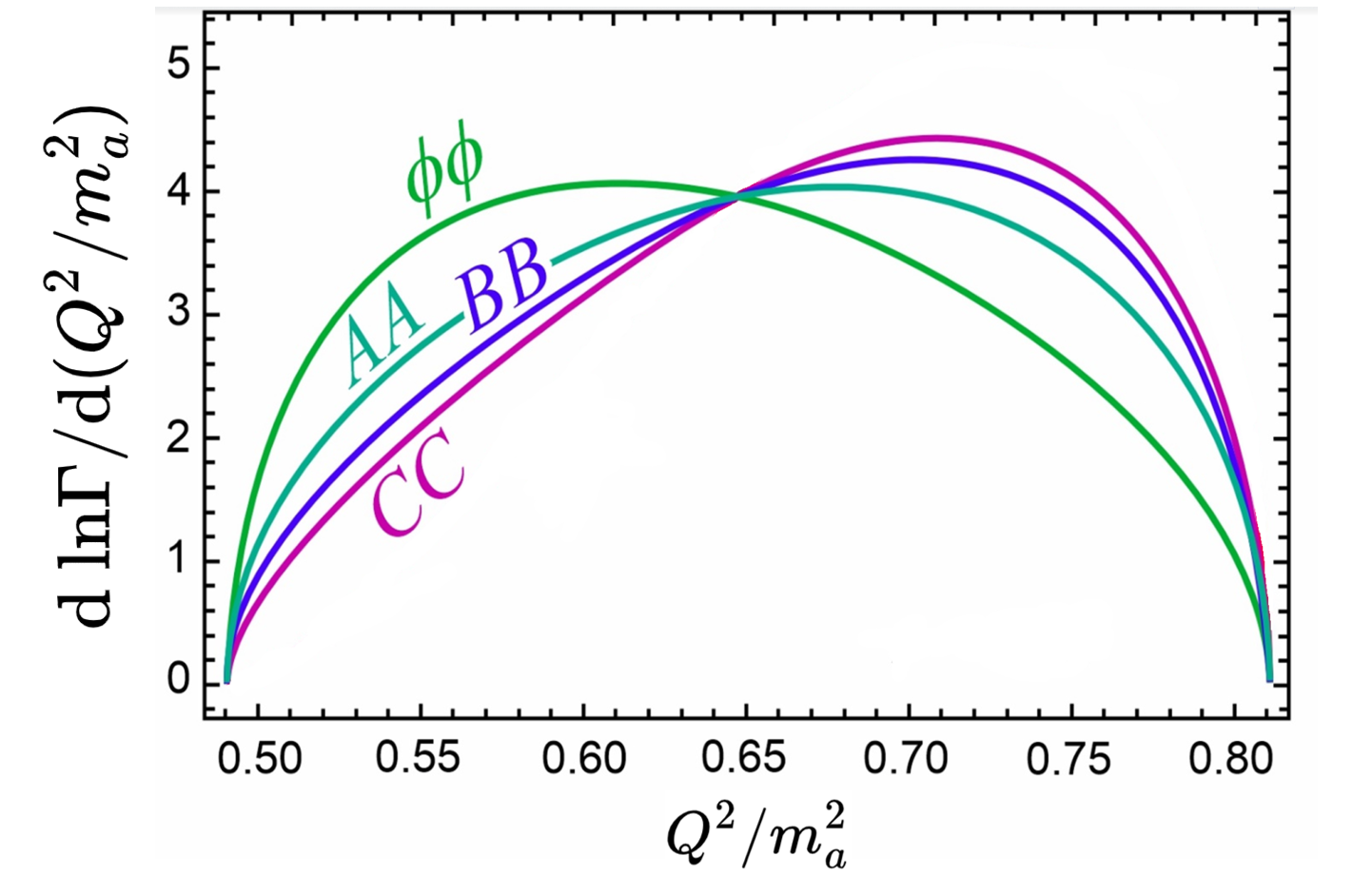}
         \captionof{figure}{Normalized differential decay rates for $q_A\rightarrow q_BXX$ in function of the normalized squared impulsion of the pair of dark fields $Q^2/m_a^2$. $m_X$ is taken identical for all form fields. We have $m_b/m_a=0.1$, $m_x/m_a=0.35$}
    \label{fig:placeholder}

    \end{minipage}
   
\end{figure}

\section{Phenomenological implications}

To summarize, the higher-form description of a particle departs from the standard picture in the three key aspects enumerated below. We will draw a sketch of how each of these points can motivate phenomenologically the search for higher-form dark matter.

\begin{enumerate}
\item The Stueckelberg decomposition shows that different DoFs will contribute at low or high energy scales
\end{enumerate}
To enlighten how the internal distribution of the DoFs can discriminate between the standard and higher-form pictures, we look at the decay rate of a quark in a pair of dark gauge bosons, $\Gamma (q_a\rightarrow q_BXX)$ with $X=\phi,A^{\mu},B^{\mu\nu},C^{\mu\nu\rho}$. The operators used to describe the vertex are:
\begin{equation}
    \bar{\psi}_L\psi_R\phi^2,\;\;\;\;\;\;\;\bar{\psi}_L\psi_RA_{\mu}A^{\mu,}\;\;\;\;\;\;\;\bar{\psi}_L\psi_RB_{\mu\nu}B^{\mu\nu},\;\;\;\;\;\;\;\bar{\psi}_L\psi_RC_{\mu\nu\rho}C^{\mu\nu\rho}.
\end{equation}
These are all renormalizable, gauge-breaking operators sharing the same structure (thus none of them are related through duality), so the decay rate is ultimately sensitive only to the kinematics of the internal DoFs. \cref{fig:placeholder} displays the normalized decay rate in functions of the impulsion of the dark pair. The discrepancy between the production of $\phi$ and $C$ is the most striking here: whereas the pair production of $0$-form peaks at low energy, the fact that $C^{\mu\nu\rho}\sim F_B^{\mu\nu\rho}/m_X$ leads to a peak at high-energy for the pair production of $3$-form. This is specifically relevant in the quest for heavy dark-matter at colliders: a scalar particle would be expected at greater energy if embodied in a $3$-form. 
\begin{enumerate}[start=2]
\item The dominant effective interactions with matter are structurally different
\item The gauge invariant operators are structurally different
\end{enumerate}
As specified in the previous section, the interest of the second point is immediate as it allows one to naturally boost certain couplings: As duality swaps the orders of mass, operators that were suppressed in a picture are now dominant in its dual, and reciprocally. The third point allows us to go one step ahead: upon imposing dark-gauge (or shift) invariance to the Lagrangian, we can now fully discriminate between the standard and higher-form description of a particle. We will give here two examples.

Let us consider a light, spin-$0$ particle (like an ALP). In the usual, $0$-form representation, the dominant coupling is a dimension-five axion-like coupling $(\partial_{\mu}\phi)\bar{\psi}\gamma^{\mu}\gamma^5\psi$ as the Yukawa coupling to fermions is forbidden by shift-symmetry. In the $3$-form representation, however, the renormalizable, axion-like interaction $\varepsilon_{\mu\nu\rho\sigma}C^{\mu\nu\rho}\bar{\psi}\gamma^{\sigma}(\gamma^5)\psi$ is forbidden and the dominant interaction is the Yukawa-like, dimension-five operator $\varepsilon_{\mu\nu\rho\sigma}F_C^{\mu\nu\rho\sigma}\bar{\psi}\psi$.
It is also possible to discriminate between the $1$-form and $2$-form representation of a gauge-invariant dark photon: in the $2$-form case, the kinetic mixing with the SM photon $B_{\mu\nu}F^{\mu\nu}_{\gamma}$ becomes forbidden as well as the EDM/MDM interactions $B^{\mu\nu}\bar{\psi}\sigma_{\mu\nu}(\gamma^5)\psi$. The dominant interaction with fermions becomes the vector/axial dimension-five coupling $\varepsilon_{\mu\nu\rho\sigma}F_B^{\mu\nu\rho}\bar{\psi}\gamma^{\sigma}(\gamma^5)\psi$.

\section{Conclusion}

Despite the constraints imposed by duality, higher-forms make compelling and natural candidates for dark matter. They have been shown to depart from the usual scalar and vector fields whenever interactions are turned on, and the rich phenomenology that comes from it is plainly illustrated in the three following points

\begin{itemize}[itemsep=0cm]
    \item Effective operators that were considered irrelevant in the standard picture are now dominant in the dual description, and reciprocally. A reinterpretation of the current limits on ALPs and dark photons would be called for in order to account for the higher-form description.
    \item Operators that were gauge-breaking in the standard picture are gauge-preserving in the dual description, and vice-versa. This could lead to natural kinetic mixing and EDM/MDM-free dark photons, or Yukawa-coupling ALPs.
    \item Degrees of freedom do not dominate at the same energy regimes in two dual pictures. Thus, heavy higher-forms (at TeV or above) would have unique experimental signatures at colliders. 
\end{itemize}
Ultimately, higher-form fields could also provide deeper insights into our understanding of QCD and its vacuum structure. One way to relate these theoretical questions to the phenomenological outcomes presented above is to investigate the link between higher-forms and chiral anomalies, looking at the couplings of $B^{\mu\nu}$ and $C^{\mu\nu\rho}$ to fermionic triangle diagrams \cite{Plantier:2026rsv}.


\begin{thebibliography}{99}
\bibitem{Kalb:1974yc}
M.~Kalb and P.~Ramond,
\textit{Classical direct interstring action}, Phys.~Rev.~D~9 (Apr, 1974) 2273-2284.

\bibitem{Curtright:1980yk}
T.~Curtright, \textit{Generalized Gauge Fields}, Phys.~Lett.~B 165 (1985) 304–308.
\bibitem{Hell:2021wzm}
A.~Hell, \textit{On the duality of massive Kalb-Ramond and Proca fields}, JCAP 01 (2022), 01
056, [2109.05030].
\bibitem{Plantier:2025hcm}
C.~Plantier and C.~Smith, \textit{Dark higher-form portals and duality}, Phys.~Rev.~D 112 (2025),
no. 7 075043, [2506.04795].
\bibitem{Stueckelberg:1938hvi}
E.~C.~G.~Stueckelberg, \textit{Interaction energy in electrodynamics and in the field theory of
nuclear forces}, Helv. Phys. Acta 11 (1938) 225–244.
3265–3348, [hep-th/0304245].
\bibitem{dvali2022strongcpgravity}
G.~Dvali, \textit{Strong-cp with and without gravity}, e-print (2022) [2209.14219].
\bibitem{Polchinski:1998rr}
J.~Polchinski, String theory. Vol. 2: Superstring theory and beyond. Cambridge
Monographs on Mathematical Physics. Cambridge University Press, 12, 2007.
\bibitem{Plantier:2026rsv}
C.~Plantier and C~Smith, \textit{Dark higher-form fields and triangle anomalies}, JHEP 05 (2026) 029, [2602.14839] .
\end{thebibliography}
\end{document}